\documentclass[twocolumn]{elsart3p}
\usepackage{epsfig}

\begin{document}

\begin{frontmatter}



%\title{Low energy neutron resoponse of PARIS phoswich (LaBr$_3$-NaI) detector}
\title{Study evolution of fragment energy spectrum in compound and elemental absorber with thickness via effective charge correction}


\author[label1,label2]{Rajkumar Santra},
\author[label3]{V.G.Vamaravalli},
\author[label4]{Ankur Roy}
\author[label1]{Balaram Dey}
\author[label1]{Subinit Roy\corauthref{cor}},
\corauth[cor]{Corresponding author.}
\ead{subinit.roy@saha.ac.in}

\address[label1]{Saha Institute of Nuclear Physics, 1/AF Bidhan Nagar, Kolkata-700064, India}
\address[label2]{Homi Bhabha National Institute, Anushaktinagar, Mumbai-400094, India}
\address[label3]{Department of Physics, Andhra University, Visakhapatnam, India}
\address[label4]{Department of Physics, Jadavpur University, Kolkata - 700032, India}

\begin{abstract}
The energy loss behaviour of fission fragments (FF) from $^{252}$Cf(sf) in thin Mylar ($H_8 C_{10} O_4$) and Aluminium 
absorber foils have been revisited. The aim is to investigate the observed change in the well known asymmetric energy of 
spontaneous fission of $^{252}$Cf as the fragments pass through increasingly thick absorber foils. Two different types 
of absorbers have been used- one elemental and the outher an organic compound. 
%Evolution of energy spectrum of fission fragments from $^{252}$Cf(sf) with increasing thickness of the two low mass absorbers having very similar charge to mass ratio has been probed. 
The stopping powers have been determined as a function of energy for three fragment mass groups with average masses with $<A>$ = 106.5, 141.8, 125.8 corresponding to light, heavy and symmetric fragment of $^{252}$Cf. 
%The energy loss data have been compared with the predictions of SRIM 2013 code. The best representations of the data have been achieved using the effective Z correction term in the stopping power relation from the classical Bohr theory. 
Using the effective charge (Z$_{eff}$) in the stopping power relation in the classical Bohr theory best describes the stopping power data.
Spectrum shape parameters, subsequently have been extracted from the energy spectra of fission fragments for different foil thickness. 
The effective charge (Z$_{eff}$) correction term determined from the stopping power data is then used in the simulation for the absorber thickness
dependence of the shape parameters of the energy spectrum. The present simulation results are compared with the TRIM prediction. The trends of the absorber thickness dependence of the spectrum shape parameters, 
for both Mylar and Aluminium are well reproduced with the present simulation.
%TRIM simulations for the absorber thickness dependence of the parameter values corroborate with the $Al$ data but not with the Mylar data. In this connection, the variation of the correction factor for effective Z with decreasing velocity of the projectile in the absorbers of elemental Aluminium and organic compound Mylar have been compared. 

\end{abstract}

\end{frontmatter}

% main text
\section{Introduction}

The energy loss mechanism of heavy charged particles, {\it e.g.} the fission fragments, unlike the light charged particles, 
is much more complicated due to the variation of effective charge of the heavy particles as the velocity of the 
particles changes in the medium \cite {knipp}. Also towards the end of the flight path, the energetic heavy particle loses 
major part of its energy in atom-atom collisions. Theories of specific energy loss of charged particles in matter have been 
presented in seminal works of Bohr \cite{bohr1,bohr2} and Lindhard, {\it et al.} \cite{Lindhard}. 

%The purpose of the present work is to 
%investigate the evolution of FF energy spectrum with increasing the absorber thickness in two different solid medium, 
%{\it viz.}, metallic Aluminium and organic polymer Mylar ($C_{10}H_{8}O_{5}$).

The measurement of specific energy loss of heavy ions in different media have also been performed extensively in the past and compared 
with various theoretical models \cite{hpaul,shafrir,pickering,laichter,govil,Moak}. Of different heavy projectiles, fragments
from asymmetric fission of $^{252}$Cf nucleus provides an interesting domain for study of energy loss of heavy charged particles. 
The decay fragments are neutron rich, heavy in nature with velocities in the range of 0.4 $\le$ $E/A$ $\le$ 1.3 MeV/u. As $^{252}$Cf undergoes 
asymmetric fission, one gets two distinct groups of heavy charged particles for a simultaneous study of energy loss behaviour 
in a medium. In a recent work, Biswas, {\it et al.} \cite{biswas1} have observed that the {\it shapes} of energy or velocity spectra of the two 
fragment groups change distinctly as the thickness of absorber medium is increased. In their case the absorber medium was Mylar, an organic
polymer.

In the present study, we investigated the distinctive change in the energy spectra of the two groups of fission fragments from $^{252}$Cf in two different absorber media.

%energy loss behaviour of fission fragments from $^{252}$Cf in two different absorber media, in organic polymer Mylar ($C_{10}H_{8}O_{5}$) and in metallic Aluminium, as the thickness of the absorber is increased. 
We used the organic compound Mylar along with the elemental Aluminium as the two types of absorber media.

Mylar with its covalent bond structure for the electrons has less number of free electrons compared to Aluminium, which 
approaches the ideal metallic condition of free valence electron gas surrounding the array of metallic ions. The specific 
energy loss for most probable {\it light}, {\it heavy} and {\it symmetric} fragments in Mylar and Aluminium absorber foils of varying thickness have been measured.

The semi-empirical fit using Bethe-Bloch formula with dynamic effective charge correction is used to describe data and the parameters of an enparical expression for effective charge, Z$_{eff}$, have been derived.
Finally, we provided the explanation for the evolution of shapes of the energy distributions of light and heavy fragment groups in two different absorber media including the dynamic effective charge correction factor in Bethe-Bloch formula.

%In the next step, we explained the observed change in the shape of the energy distributions of light and heavy fragment groups, for these two absorber media, by including the extracted effective charge correction factor with decreaing velocity in the 
%Bethe-Bloch formula for stopping power.

\begin{figure}[h]
\begin{center}
\includegraphics[scale=0.22]{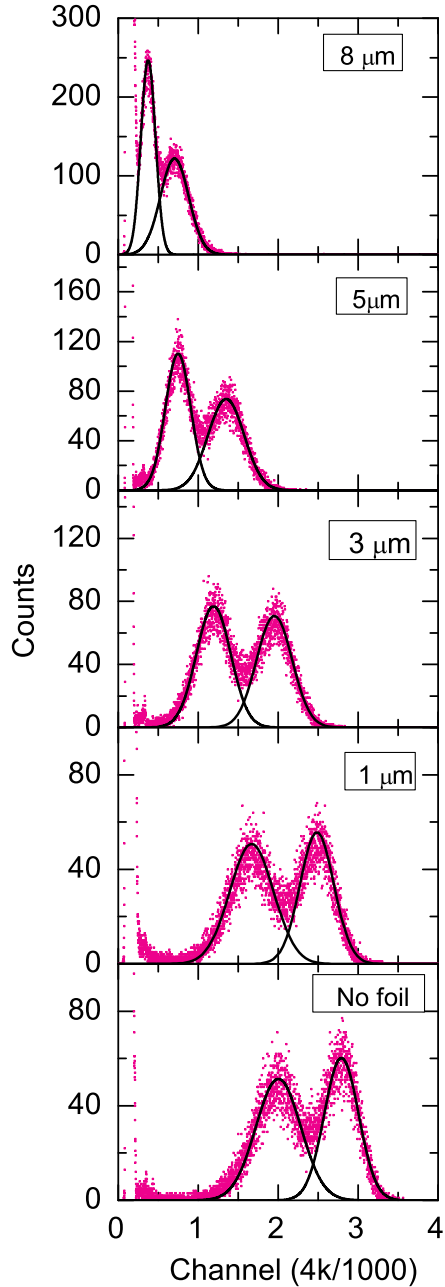}
\caption{\label{fig1} \small \sl Energy spectrum of fission fragments without foil and with Mylar foils of different thickness.}
\end{center}
\end{figure}

\begin{figure}[h]
\begin{center}
\includegraphics[scale=0.22]{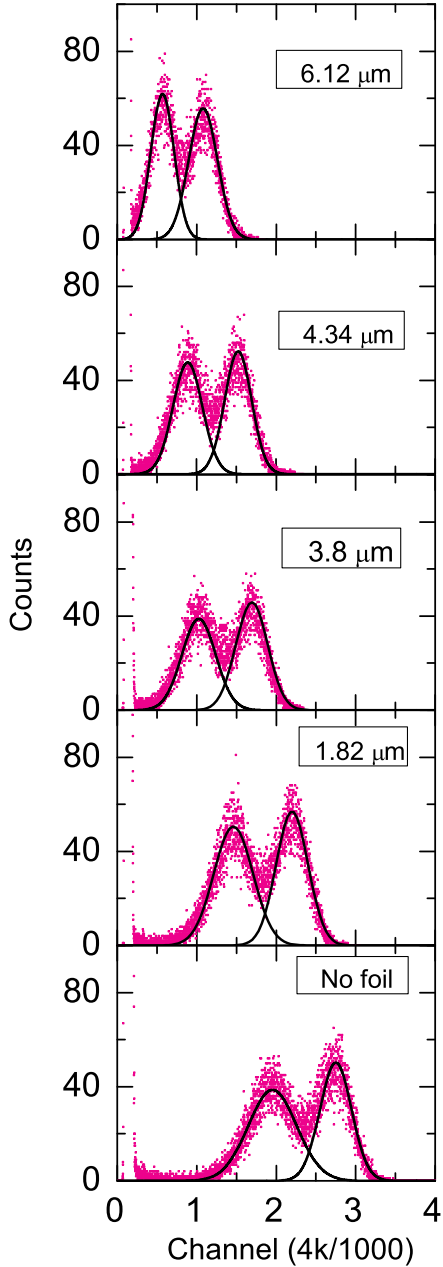}
\caption{\label{fig2} \small \sl Energy spectrum of fission fragments without foil and with Aluminium foils of different thickness.}
\end{center}
\end{figure}

\section{Experimental set-up}

The setup consists of a vacuum chamber with a source holder and a detector mount. The absorber foil can be placed in between 
the source holder and the detector mount. The chamber is then evacuated and maintained at a pressure $ \sim 10^{-3}$ Torr during the measurements. 
An ORTEC Silicon Surface Barrier detector (Model No: BU-015-200-100, Serial No: 3.3-271D.5) of effective thickness of 100 $\mu$m at an operating 
bias of 50 volt is used in the experiment. Normal electronics setup has been adopted for processing the signal.The signal is digitized
through NIM based Dual ADC in MPA3 data acquisition system from FASTCOMTEC and stored in the desktop for subsequent analysis.

\subsection{Comparison of FF pulse height spectra in Mylar and Aluminium absorbers} 

In this work we studied the relative change in energy distributions of light and heavy groups of fission fragments with increasing 
thickness of the absorber foil. Two different types of absorbers - organic compound Mylar and metallic Aluminum having very similar 
$\frac{Z_2}{A_2}$ ($\sim 0.5$)- have been used. An interesting 
behavior is observed in the study of evolution of the energy spectrum of the FF as the foil thickness is increased. The nature of evolution 
of the energy distribution of fission fragments from $^{252}$Cf with foil thickness is distinctly different for Mylar and Aluminium. The variation
with increasing thickness of Mylar and Aluminium foils are shown in Figs. \ref{fig1} and \ref{fig2},

To further investigate the changing nature, we fitted both heavy and light FF energy spectra by Gaussian functions and the ratio of area under the 
light and the heavy fragment peaks have been estimated. It is observed that the ratio remains within a range of 10\% relative to the 
value obtained without any foil. This ensures that while estimating the spectrum parameters no counts under the peaks have been missed.

\subsection{Energy calibration of silicon detector for FF}

The suitability of the detector used for FF detection is established following the prescription of Schmitt and Pleasonton 
\cite {shafrir, schmitt1, schmitt2}. The resulting spectrum shape parameters corroborate nicely with the expected values
of Schmitt and Pleasonton\cite {shafrir, schmitt1, schmitt2} that establishing the goodness of the detector for the energy
loss measurements of the fission fragments.

In the next step, we used the mass-dependent energy calibration equation from Ref. \cite{knoll} to obtain the fragment energy in terms of 
its mass ($m$) and the corresponding pulse height ($x$) from the silicon surface barrier detector. The relation is  

\begin{equation}
\label{eqn1}
E(x,m) = (a+ a^{\prime}m)x + b + b^{\prime}m 
\end{equation}
  
The constants in Eq.\ref{eqn1} depend on the spectrum shape parameters and the required expressions are given as in Ref. \cite{Wiss}
The values of the constants of mass dependent calibration equation for the present experiment have been shown in Table \ref{tab3}.

\begin{table}[!ht]
\caption{\label{tab3} Estimated values of calibration constants.}
\vspace{5mm}
%\begin{ruledtabular}
\begin{center}
\begin{tabular}{|c|c|c|c|}
\hline 
 ~~ a~~ & ~~ a$^{\prime}$ ~~& ~~  b ~~&   b$^{\prime}$~~ \\ \hline
&&& \\ 
~~ 0.0299      ~~   & ~~ 3.4x10$^{- 5}$ ~~& ~~ 6.38 ~~& ~~ 0.0172 ~~\\
&&& \\ \hline 
\end{tabular}
%\end{ruledtabular}
\end{center}
\end{table}

\begin{table*}[!ht]
\caption{\label{tab4} The energy and PHD values of light, heavy and symmetric FF.}
\vspace{5mm}
%\begin{ruledtabular}
\begin{center}
\begin{tabular}{|c|c|c|c|c|c|c|}
\hline 
System &       \multicolumn{2}{l}{Energy E(x,m) in MeV}         &   & \multicolumn{2}{l}{PHD in MeV} &\\ \hline 
 FF           & Light                 & Heavy                  & symmetric  & Light & Heavy & symmetric \\ \hline 
Present study & 102.56$\pm$ 1.0       & 78.48 $\pm$ 0.62       & 91.76 $\pm$ 0.8  & 9.2   & 12.5  & 10.76 \\ \hline
From          & 103.0 $\pm$ 0.6 \cite{Knya}    & 78.9 $\pm$ 0.5 \cite{Knya}       &            & 16 \cite{bozor,finch}        & 17 \cite{bozor, finch} &\\ 
reference     & 103.0 $\pm$ 0.5 \cite{schmitt2}    & 79.37 $\pm$0.5 \cite{schmitt2}       &    & 13 \cite{kone}        & 14 \cite {kone} &\\ 
              & 102.54 $\pm$ 0.94 \cite{schmitt1} & 78.68 $\pm$ 0.5 \cite{schmitt1}     &            &          &  &\\ \hline
\end{tabular}
%\end{ruledtabular}
\end{center}
\end{table*}

Using calibration Eq. \ref{eqn1} with the constants given in Table \ref{tab3}, the energies of light, heavy and symmetric fragments are estimated. The 
average mass values of the fragments are taken from Ref. \cite{Knya}. The energy values of respective fragments are given in Table \ref{tab4}. The estimated 
values of the fragment energies, without any absorber, compare well with those reported in the literature. In Table \ref{tab4}, we have also presented our estimation 
of pulse height defect (PHD) of each fragment. The pulse height defect (PHD) has been evaluated as a difference of expected energy from alpha calibration and the actual 
observed energy for respective fragments. A comparison of the resultant PHD values for the light and heavy fragments are shown in Column 4 of the table. The energy 
calibration scheme is then followed in subsequent determination of energy loss of fragments in the absorber foil.

\section{FF energy loss in Mylar and Aluminum absorbers}
It is difficult to measure the energy loss of each and every FF mass from the energy spectrum. We measured instead the energy loss of most probable complimentary 
light and heavy fragments and that of the symmetric fragments using the mass dependent energy calibration equation of FF for the peak locations of {\it light}, 
{\it heavy} and {\it symmetric} fragments for different thicknesses of the absorbers. 
Energy loss is measured as the difference of two energies {\it viz.} $E_{foil}$, the fragment energy after passing through the selected foil and $E_{hole}$  
the fragment energy without any foil. The specific energy loss is then estimated at an effective particle energy $E$ within the thickness $\Delta X$ as
\begin{eqnarray}
\label{eqn3} 
\frac{\Delta E}{\Delta X} (E) = \frac{E_{hole}-E_{foil}}{h}
\end{eqnarray}
where ${h}$ is the thickness of absorbing foil and $E$ is the effective particle energy. 
For thin absorber ($h$ less than $\approx$ 1$\mu$m) with $\Delta E \ll E$, the approximation $ E = (E_{hole} + E_{foil})/{2}$ is used to estimate
the effective energy in the present work. 
But for higher target thickness where $\Delta E \ll E$ does not hold, we used a numerical approach with finite foil thickness correction to 
estimate the effective E following the expression given in Ref. \cite{mertens}.

\begin{figure}[h]
\begin{center}
\includegraphics[scale=0.25]{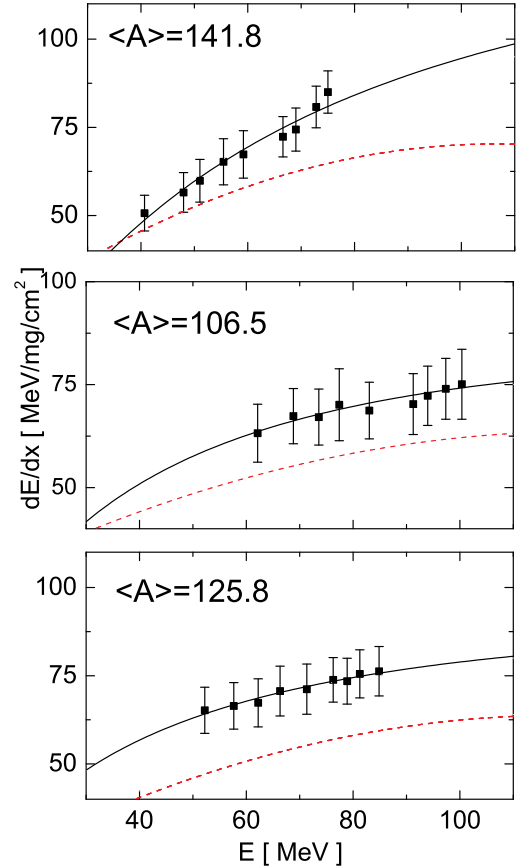}
\caption{\label{fig3} \small \sl  Stopping power data with Mylar foils, for heavy  FF with $<$A$>$=141.8 (top panel), 
light FF with $<$A$>$=106.5 (middle panel) and symmetric FF with $<$A$>$=125.8 (bottom panel). Solid square points with error bar 
are the present experimental data. Solid black lines show present semi-empirical fits and red dashed lines denote calculation with SRIM 2013 code.}
\end{center}
\end{figure}

\begin{figure}[h]
\begin{center}
\includegraphics[scale=0.25]{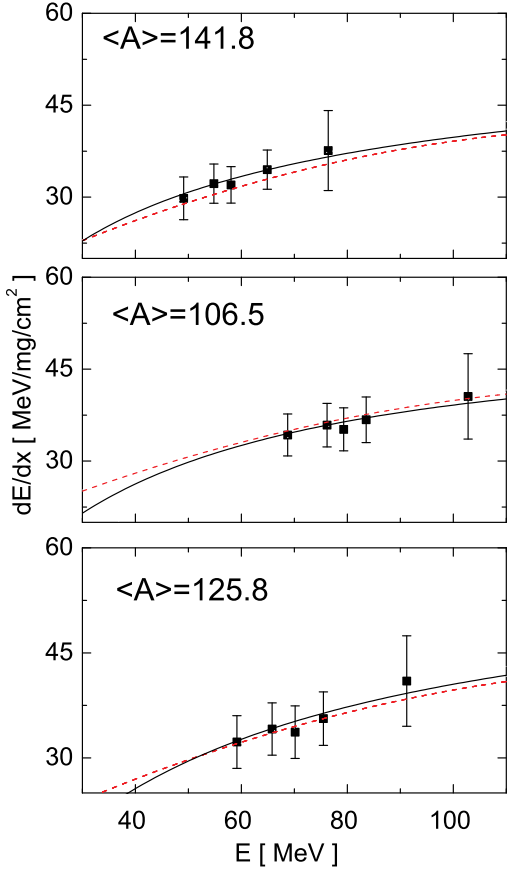}
\caption{\label{fig4} \small \sl Stopping power data with Aluminium foils, for heavy FF with $<$A$>$=141.8 ( top panel), light FF 
with $<$A$>$=106.5 (middle panel) and symmetric FF with $<$A$>$=125.8 (bottom panel). Solid square points with error bar are the 
present experimental data. Solid black lines show the present semi-empirical fit and red dashed lines give the calculation with SRIM(no effective Z) 
2013 code.}
\end{center}
\end{figure}

The estimated stopping power values as function of incident energy are plotted in Figs. \ref{fig3} and \ref{fig4}. The plots are shown for
three different average mass values corresponding to the two peaks and the minimum in the mass spectrum.

\subsection{Estimation of uncertainty}
In the determination of uncertainty associated with the measured specific energy loss, we used the standard error propagation technique. 
The uncertainty in specific energy loss primarily comes from the uncertainties in energy loss $\Delta E$ of the fragments in the absorber and in the 
determination of foil thickness. Taking into consideration all the factors the uncertainty of $\Delta E$ varies from 0.6 to 1.9 \% and 
the accuracy of foil thickness is typically in the range of 10 to 16\%. The thickness of each absorber foil is determined separately by the $\alpha$-energy 
loss technique, using a 3-line $\alpha$-source. 
The overall uncertainty of the specific energy loss value in the present measurement is primarily determined by the uncertainty in the foil thickness and is 
within 10-17\%. The error resulting from the use of the averaged mass and atomic number instead of the actual values, estimated using the code 
SRIM 2013, is small compared to the overall uncertainty mentioned above. Hence this contribution has not been considered.
The errors are shown in Figs \ref{fig3} and \ref{fig4}.

\section{Semi-empirical description of FF energy loss}

%In Figs. \ref{fig3} and \ref{fig4}, two different curves have also been shown in comparison with the data. The dashed curves are the predictions of 
%SRIM 2013 code. In case of metallic Aluminium absorber, SRIM could describe the variation of effective stopping power as a function of energy relatively well. 
%But for organic polymer Mylar, the predictions of SRIM are systematically lower than the data over all energies. The same trend has also been observed 
%by Knyazheva, {\it et al.} \cite{Knya}.

%The solid curves have been generated after the incorporation of the variation of effective charge of the ion as a function of reducing velocity in the medium while estimating the electronic stopping power.

According to classical theory of Bohr \cite{bohr2}, electronic stopping power of an ion with charge $Z_1$ moving with velocity $v$ in a medium 
consisting of atoms having charge and mass $Z_2$, $A_2$, is given as 
\begin{eqnarray}
\label{eqn4} 
-\frac{dE}{dx} = 3.07 \times 10^4 \frac {Z_2}{A_2} \frac {Z_1^2}{\beta^2} \ln(\frac {m_e v^2}{I})
\end{eqnarray}
in the non-relativistic limit and in the unit of $MeV/mgcm^{-2}$. Here $\beta = v/c$, c is the speed of light. The mean ionization and 
excitation potential $I = I_0 Z_2$ with $I_0 \approx$ 10 eV. Now fast moving heavy ions, depending on their velocities, can pick up electrons from or lose 
electrons into the medium. Thus the effective ionic charge fluctuates around a certain equilibrium value of $Z_{eff}$ \cite{bohr1}. The resulting screening 
of Coulomb field of the nuclear charge affects the stopping power. The effective charge is expressed as $ Z_{eff} = \gamma Z_1 $ where multiplicative factor $\gamma$ 
is the effective charge parameter which has a complicated dependence on the atomic number $Z_2$ of absorber foil and the energy or velocity of the fragment
moving through the absorber medium. The general semi-empirical form for $\gamma$, as suggested by Bohr, can be written as \cite{north, biswas2}.
\begin{eqnarray}
\label{eqn6} 
\gamma = 1-a_0 \exp(-a_1 \frac{v}{v_0 Z_2^{2/3}})
\end{eqnarray} 
where $a_0$ and $a_1$ are constants and $v_0$ is the Bohr velocity.
We obtained three sets of values for $a_0$ and $a_1$ each by fitting the stopping power data in Mylar and in Aluminum 
foils is shown in Figs \ref{fig3} and \ref{fig4} for light, heavy and symmetric FF to construct the effective charge parameter $\gamma$. 
Since Mylar is an organic compound, we used the average value of charge and mass number for this material ($Z_2 = 4.55$, $A_2 = 9.09$)
\cite {Knya}. The fitted parameters $a_0$, $a_1$ are listed in Table \ref{tab6}. The soild lines in the figures are the fits using the semi-empirical expression. 
The dashed lines represent the prediction of SRIM-2013 without the correction for effective Z.

It is obvious from Figs \ref{fig3} and \ref{fig4} that the required correction for effective charge is significant in the covalent organic absorber Mylar compared to metallic Aluminium.
The difference in reflected in the values of parameter a$_1$ in the exponent of the relation in Eq. 4.

\begin{table}[!ht]
\caption{\label{tab6} The semi-empirical fitting parameters.}
\vspace{5mm}
%\begin{ruledtabular}
\begin{center}
\begin{tabular}{|c|c|c|c|c|}
\hline 
Foil  &  FF group & $a_0$  &  $a_1$  & Reduced $ \chi^2$\\ \hline
          & Light    & 1.11   & 3.56 & 1.28  \\ 
Mylar     & Heavy    & 1.15   & 3.54 & 2.7  \\ 
          & Symmetric & 1.06   & 3.13 & 2.78   \\ \hline
          & Light    & 1.05   & 5.81 & 0.57  \\ 
Al        & Heavy    & 1.00   & 3.98 & 0.74  \\ 
          & Symmetric & 1.04   & 5.10  & 1.24   \\ \hline
\end{tabular}
%\end{ruledtabular}
\end{center}
\end{table}

\section{Results and Discussion}
\begin{figure}[h]
\begin{center}
\includegraphics[scale=0.25]{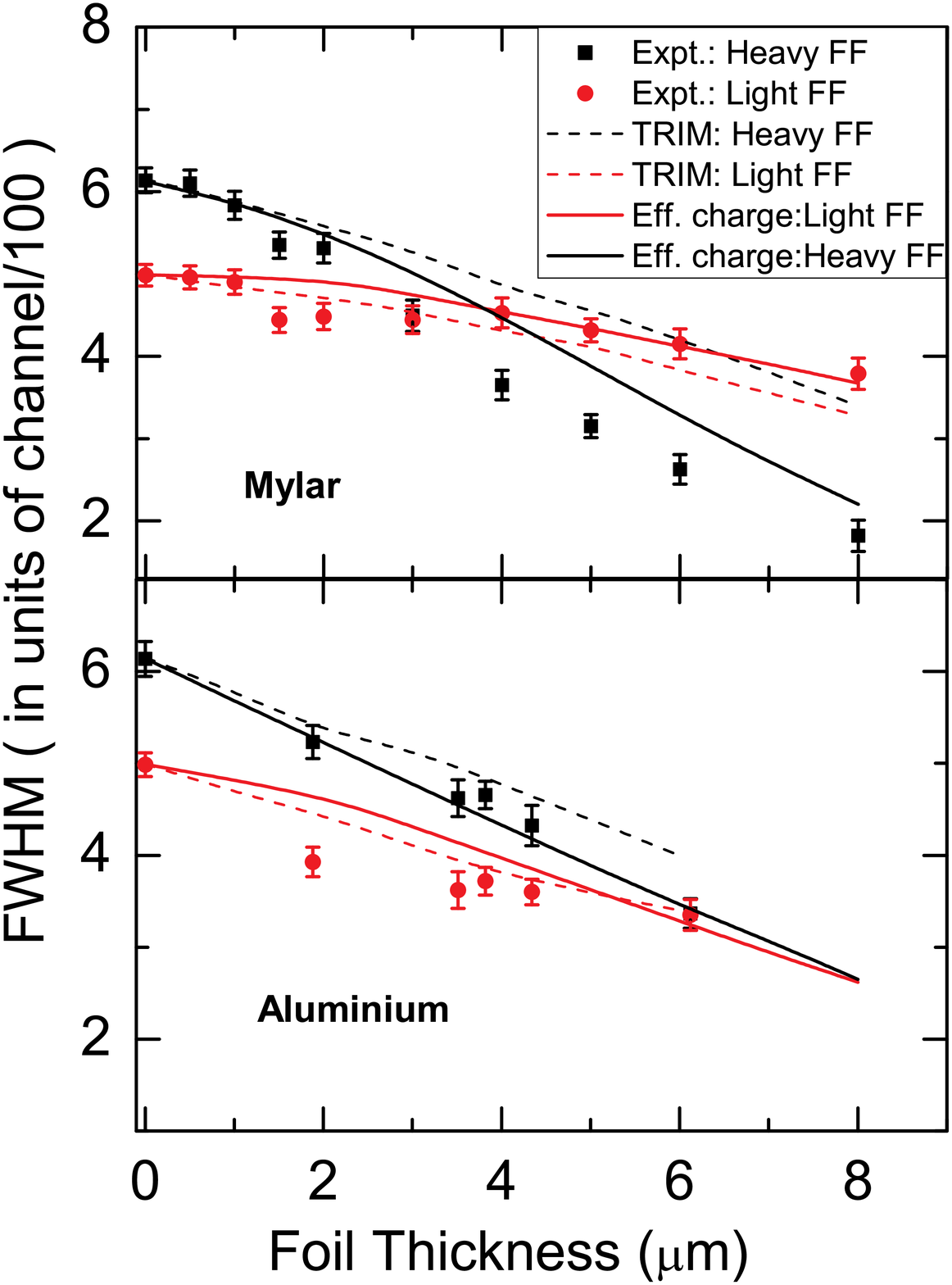}
\caption{\label{fig7} \small \sl Variation of FWHM of light- and heavy-fragment energy peaks with increasing thickness of absorber foil.}
\end{center}
\end{figure}

\begin{figure}[h]
\begin{center}
\includegraphics[scale=0.25]{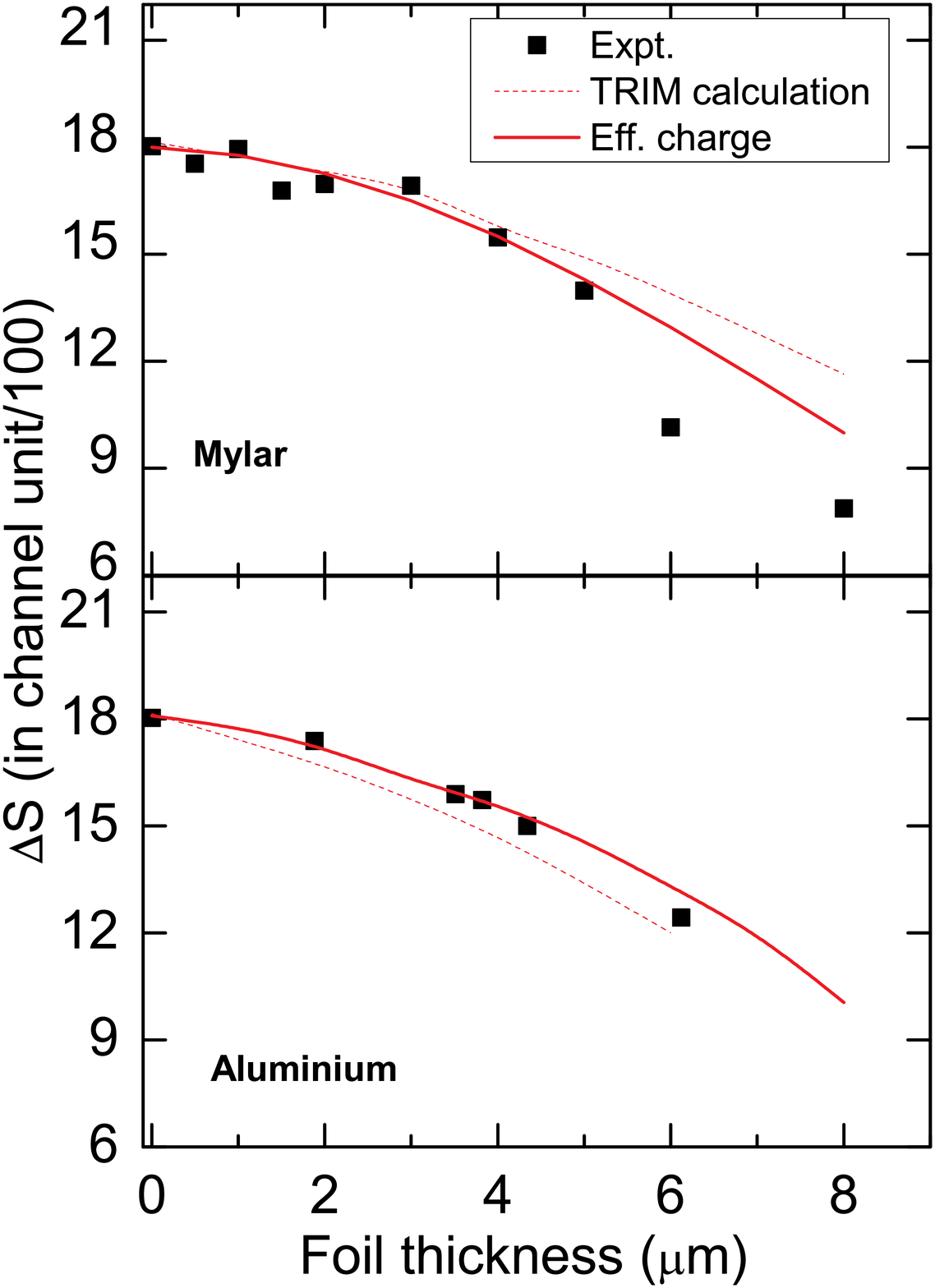}
\caption{\label{fig8} \small \sl Variation of FWTM or $\Delta$S of light- and heavy-fragments with increasing thickness of absorber foil.}
\end{center}
\end{figure}

As mentioned earlier, in the current work, we studied the relative change in energy distribution patterns of light and heavy groups of fission fragments with increasing 
thickness of the absorber foil. An interesting 
behavior is observed in the study of evolution of the energy spectrum of the FF as the absorber foil thickness is increased. The nature of evolution 
of the energy distribution of fission fragments from $^{252}$Cf with foil thickness is distinctly different for Mylar and Aluminum though having very similar $\frac{Z_2}{A_2} \approx 0.5$. The variation
with increasing thickness of Mylar and Aluminum foils are shown in Figs. \ref{fig7} and \ref{fig8}, respectively. To further investigate the 
changing nature, we fitted both heavy and light FF energy spectra by Gaussian functions and the ratio of area under the light and the heavy 
fragment peaks have been estimated. It is observed that the ratio remains within a range of 10\% relative to the 
value obtained without any foil. This ensures that while estimating the spectrum parameters no counts under the peaks have been missed.

%In Fig. \ref{fig7}, we have shown the variation of the ratio value for different foil thicknesses in case of Mylar absorber.
%\begin{figure}[h]
%\begin{center}
%%\includegraphics[scale=0.30]{figure7.eps}
%\includegraphics[scale=0.30]{area_ratio_mylar_al.eps}
%\caption{\label{fig7} \small \sl Ratio of area under the light and heavy fragment energy peaks for different Mylar foil thicknesses.}
%\end{center}
%\end{figure}

Three different spectrum shape parameters, viz., the FWHM values of the fitted energy spectra of two individual FF groups and the parameter $\Delta$S that 
gives the width of the total energy spectrum estimated at the FWTM level, have been used to characterize the spectrum. 
The plots of FWHM and $\Delta$S vs. foil thickness with Mylar and Aluminum foils are shown 
in Figs. \ref{fig7} and \ref{fig8}, respectively. The deviation from the TRIM-2013 simulated curve (dashed carves) at higher foil thickness is clearly visible in the 
figures. But semi-empirical description including dynamic effective charge better reproduce the experimental features for both the absorber media.
Interestingly, the FWHM of heavy fragment peak falls off much more sharply compared to the TRIM prediction with increasing thickness
of Mylar foil, while the FWHM of the light fragment peak decreases slowly and matches with the TRIM prediction at larger foil thickness region.
In case of Aluminum absorber, the FWHM-s of light and heavy fragment peaks follow the trend yielded by TRIM simulation.
The parameter $\Delta$S describing the overall width of the $^{252}$Cf fission fragment energy spectrum in Mylar shows a much steeper fall
relative to the simulation data as the foil thickness increases. On the other hand in Aluminum medium, both the extracted parameter $\Delta$S 
and its simulated values show a very similar fall off with increasing foil thickness. 
%That confirms the effect of effective charge corrections is more in compound Mylar absorber than elemental Al foils.  

\section{Conclusion}
The observed behavior of the shape parameters in case of Mylar and Aluminum absorbers led us to look for the dependence of the dynamical 
quantity $\gamma$, which indicates the evolution of effective $Z$ value of the particles as they pass through the absorber medium (or as the 
velocity of the particles decrease), on the foil thickness. The resultant screening of the Coulomb field of the heavy charged nucleus of the
projectile strongly affects the energy loss behavior. Over the same thickness range, the 
neutralization of the heavy charged projectile is faster in case of organic compound Mylar, although the free electron density is lower in 
the Mylar medium compare to metallic Aluminum. Also in Mylar the rate of neutralization of heavier FF 
is quicker than the lighter FF. This is not the behavior in elemental Aluminum absorbers. In Aluminum, the energy loss behaviors of the two fragment groups are similar except for this absorber thicknesses. 
Thus energy loss mechanism of heavier fragments is essentially through atom-atom collision at low velocities in Mylar.

Energy loss behavior of fission fragments in elemental Aluminum and in organic compound Mylar has been investigated. It is observed
that the measured stopping cross section data is significantly higher compared to SRIM 2013 predictions in lighter organic medium.
The data in case of Al-absorber is quite well reproduced. The variation of the shape parameters of the spectrum with increasing
absorber thickness has been compared with results of TRIM simulation in SRIM 2013 code system. It is found that the simulated dependence of the 
FWHM of energy spectrum, on absorber thickness for heavy FF in lighter absorber over predicts the data as the foil thickness increases. However, 
for lighter FF the calculation does reproduce the data in Mylar. Over prediction is also observed for FWTM data as a function of foil thickness
in Mylar. The data for Al-absorber is well reproduced by TRIM simulation. It would be interesting to extend the investigation of energy
loss of fission fragments in elemental Beryllium foils with Z=4 for comparison with organic compound Mylar having average Z=4.5. \\

\section*{Acknowledgment}

We thank Prof. Maitreyee Saha Sarkar, Nuclear Physics Division, Saha Institute of Nuclear Physics, Kolkata for her useful suggestions 
and constant encouragement in the work.

\end{document}